\documentclass[aip,jcp,reprint]{revtex4-1}
\usepackage{amsmath}
\usepackage{graphicx}
\usepackage{verbatim}
\usepackage{color}
\usepackage{subfigure}

\begin{document}

\title{Structure of Magnetic Lanthanide Clusters from Far-IR Spectroscopy: Tb$_n^+$ ($n=5-9$)}
\author{John Bowlan}
\email{jbowlan@fhi-berlin.mpg.de}
\author{Dan J. Harding}
\affiliation{Fritz-Haber-Institut der Max-Planck-Gesellschaft, Faradayweg 4-6
14195 Berlin, Germany}
\author{Jeroen Jalink}
\author{Andrei Kirilyuk}
\affiliation{Radboud University Nijmegen, Institute for Molecules and Materials,
Heyendaalseweg 135, 6525 AJ Nijmegen, The Netherlands}
\author{Gerard Meijer}
\author{Andr\'{e} Fielicke}
\affiliation{Fritz-Haber-Institut
der Max-Planck-Gesellschaft, Faradayweg 4-6 14195 Berlin, Germany}

\begin{abstract}
Small lanthanide clusters have interesting magnetic properties, but
their structures are unknown.  We have identified the structures of
small terbium cluster cations Tb$_{n}^{+}$ ($n=5-9$) in the gas
phase, by analysis of their vibrational spectra.  The spectra have
been measured \emph{via} IR multiple photon dissociation of their
complexes with Ar atoms in the $50-250$ cm$^{-1}$ range with
an infrared free electron laser. Density functional theory
calculations using a $4f$-in-core effective core potential (ECP)
accurately reproduce the experimental far-IR spectra.  The ECP
corresponds to a $4f^{8}5d^{1}6s^{2}$ trivalent
configuration of terbium. The assigned structures are
similar to those observed in several other transition metal systems.
From this, we conclude that the bonding in Tb clusters is through
the interactions between the $5d$ and $6s$ electrons, and that the
$4f$ electrons have only an indirect effect on the cluster
structures.
\end{abstract}

\maketitle


Small clusters of lanthanide elements are promising materials for molecular
magnets, due to their large magnetic moments, and strong spin-orbit interaction.\cite{gatteschi2011anisotropic, *ishikawa2005quantum, *lin2008dinuclear,
*lin2009polynuclear}  The magnetism of lanthanide metal clusters originates
from the indirect (RKKY) exchange interaction, and the magnetic anisotropy is
due to the interaction between the asymmetric $4f$ charge cloud and the local
crystal field.  Both of these properties are highly dependent on the geometric
structure, and the lack of knowledge of the cluster structures is a major
barrier to understanding their magnetism.  To obtain fundamental insights into
their properties, the clusters can be studied in gas-phase molecular beam
experiments, which provide an environment free of external influences.

The magnetism of lanthanide clusters has been studied experimentally using the
Stern-Gerlach deflection technique. \cite{douglass1992magic, *gerion1999high,
*bowlan2010size, *van2010effect} These experiments demonstrated that lanthanide
clusters \footnote{Gd, Tb, Dy, Pr, Ho, and Tm have been studied} have large
size-dependent magnetic moments and in some cases, Curie temperatures much
higher than the corresponding bulk materials.  Theoretically, the Gd$_{13}$
cluster has been most extensively studied: In Ref.~\onlinecite{pappas1996spin,
*cerovski2000magnetization},  the cluster structures were taken to be fragments
of the bulk hcp lattice, and the magnetism was described phenomenologically with
a Heisenberg Hamiltonian. The exchange constants which describe the strength of
the interaction between local $4f$ moments due to the RKKY interaction were
taken as undetermined parameters, rather than properly estimated from electronic
structure calculations.  In fact, little to nothing is known about the physics
of the RKKY interaction in small sub-nanometer particles.  This situation stands
in sharp contrast to the present understanding of the bulk lanthanides where
many physical properties are well understood with simple models, and
first-principles theory can calculate magnetic ordering temperatures with near
quantitative accuracy. \cite{jensen1991rare, hughes2007lanthanide} The
structures of the bulk lanthanides are all known from X-ray diffraction on
single crystal and thin film samples, and the lack of such structural
information is a major barrier preventing the application of high-level theory
to small clusters.    A recent theoretical study of the Gd$_{13}$ cluster
\cite{yuan2011geometrical} which attempted to include the $4f$ electrons
concluded that the lowest energy structures were icosahedral, rather than
$hcp$-like, contradicting the previous results of. Refs.~\onlinecite{pappas1996spin,
*cerovski2000magnetization}   Despite this, lanthanide clusters have received
little or no attention from experimental structural studies.

State of the art computational electronic-structure theory still has
many difficulties with reliable prediction of cluster structures and their
related properties from first principles. This holds in particular for metals
with partially filled $d$ (transition metals) or $f$ (lanthanides, actinides)
shells or for heavy elements, like gold,  where relativistic effects are
significant. Experimental data is therefore essential for testing the validity of
predictions and in motivating methodological developments. To this end, hybrid
experimental/theoretical studies with photoelectron spectroscopy,
\cite{li2003au20, *kostko2007structure} trapped-ion electron diffraction,
\cite{xing2006structural, *johansson20082d} and far-IR spectroscopy
\cite{fielicke2004structure, harding2010probing, gruene2008structures} have
provided invaluable information on the structures of small metal clusters. 
To identify the cluster structures, we present here experimental far-IR
vibrational spectra for cationic terbium clusters in the gas phase,
mass-selectively measured by dissociation of weakly bound Ar messenger
complexes. This method has been successfully applied before, e.g., to transition
metal clusters. \cite{fielicke2007far, fielicke2004structure,
harding2010communications, harding2010probing, ratsch2005structure,
gruene2007experimental}  Terbium was chosen because there is a single naturally
occurring isotope, which is beneficial for mass spectrometric studies, and it
can be reasonably expected to behave as a prototype for the heavy lanthanides
\cite{jensen1991rare} (Gd, Tb, Dy, Ho, Er), which are all trivalent and have the
same hcp bulk lattice type.

The Free Electron Laser for Infrared eXperiments (FELIX) in Nieuwegein, the
Netherlands, was used as the far-IR source for these experiments. The
experimental setup and methods have been described in previous publications
\cite{fielicke2004structure, fielicke2007far} and we only give a brief summary
of the relevant details here. Tb clusters are produced in a laser vaporization
cluster source which is in part cooled to 77 K.  The He carrier gas is seeded
with 5 $\%$ Ar, which leads to the formation of a small population of weakly
bound Ar complexes of the Tb$_{n}^{+}$ clusters.  The cluster beam is irradiated
by the tunable far-IR radiation from FELIX and mass analyzed with a reflectron
time-of-flight mass spectrometer.  The FELIX beam used in the experiment
delivers an energy on the order of $\approx$ 10 mJ within a pulse of about 7--9
$\mu$s duration.  Mass spectra of alternating shots from the cluster source with
FELIX toggled on and off are stored into separate channels of a digitizer and in
total about 500 mass spectra are averaged per channel and frequency interval.
The IR induced relative depletion of the mass
signals for the Tb$_{n}^{+}$-Ar complexes is evaluated as a function of IR
frequency to obtain the size specific far-IR spectra.
%
%
%

\begin{figure}[]
    \centering
    \includegraphics[]{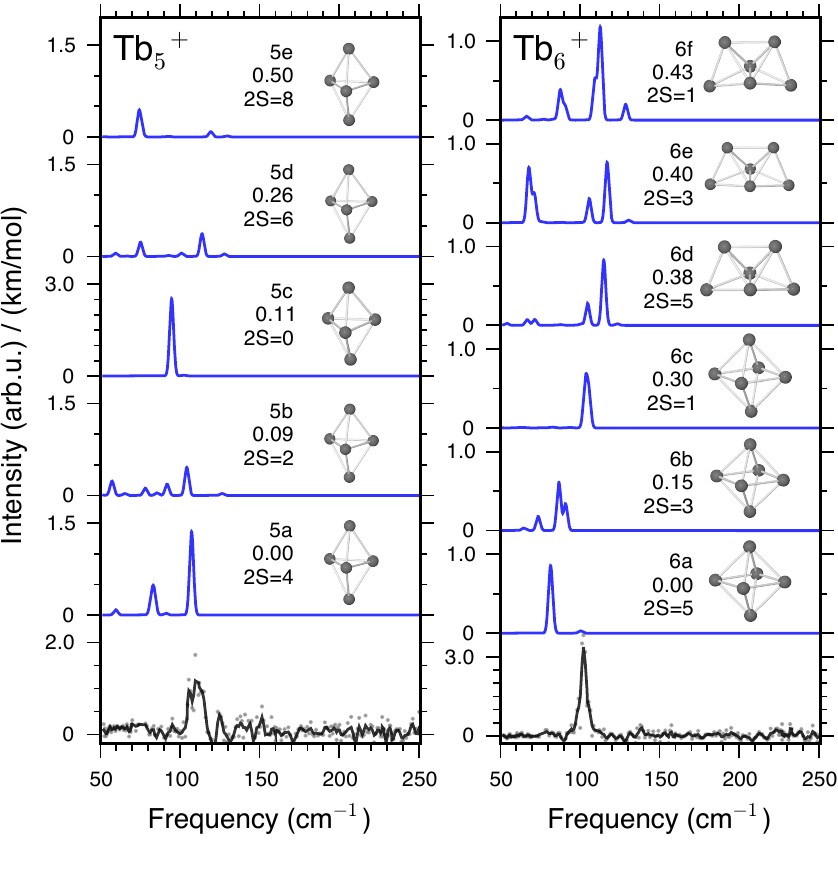}
    \caption{Experimental far-IR spectra of Tb$_{5}^{+}$ and Tb$_{6}^{+}$
    compared with the theoretically predicted spectra for a selection of
    low-lying structural isomers and $5d6s$ spin states. The experimental
    spectra are shown in relative cross sections $\sigma \propto \nu / P(\nu)
    \log( \frac{I_{\textrm{0}}}{I(\nu)})$ derived from the depletion of the Ar
    complexes, normalized by the photon fluence.  All experimental spectra are
    obtained from the complexes with a single Ar atom, except for Tb$_{5}^{+}$
    where for intensity reasons the complex with two Ar atoms has been used. To
    help the eye, the experimental raw data points (gray dots) are overlayed
    with their 5 point running average.  For each isomer its energy relative to
    the lowest energy isomer identified (in eV) and the number of unpaired
    electrons $2S$ are given.  The calculated stick spectrum is folded with a
    Gaussian line width function of 2 cm$^{-1}$ full width at half maximum.  }
    \label{fig:tb56fig}
\end{figure}

\begin{figure}[]
    \centering
    \includegraphics[]{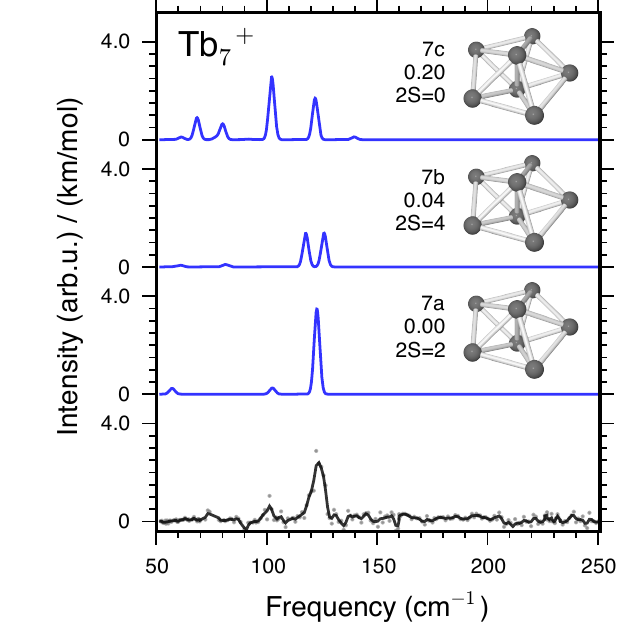}
    \caption{Far-IR spectrum and comparison to predictions for different isomers of
    Tb$_{7}^{+}$. Other plausible structures for Tb$_{7}^{+}$, such as a capped
    octahedron, or bicapped pentagonal bipyramid are either significantly higher
    in energy, or converge to a pentagonal bipyramid under local
    optimization of the total energy.}
    \label{fig:tb7fig}
\end{figure}

\begin{figure}[]
    \centering
    \includegraphics[]{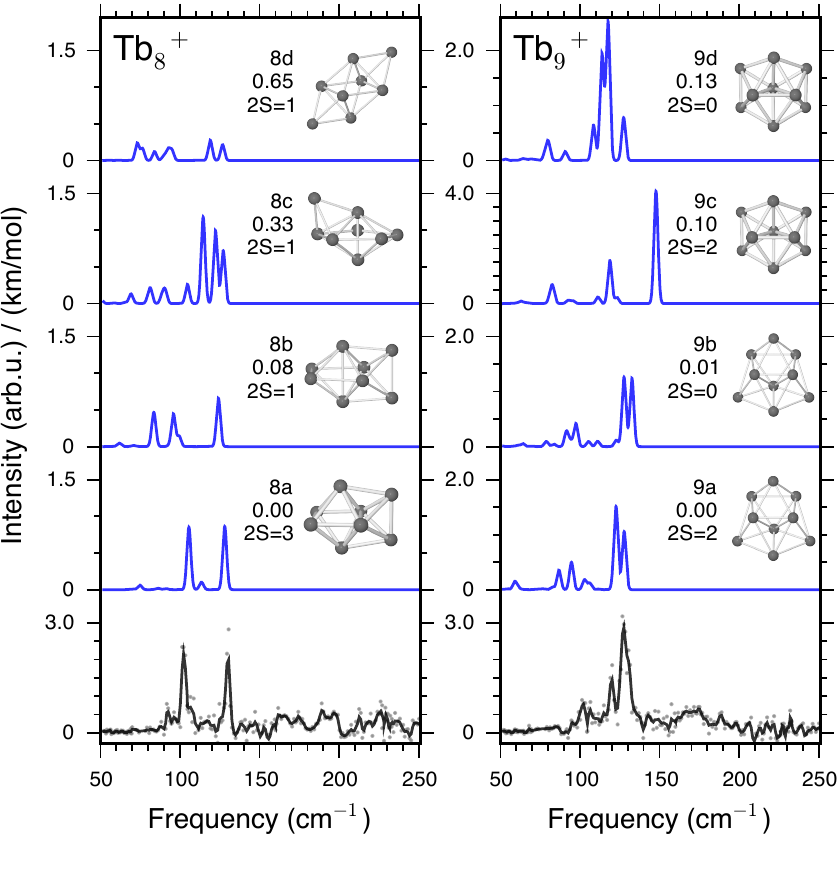}
    \caption{Far-IR spectrum and comparison to predictions for different isomers of
    Tb$_{8}^{+}$ and Tb$_{9}^{+}$.}
    \label{fig:tb89fig}
\end{figure}

The experimental IR spectra for Tb$_{n}^{+}$ ($n=5-9$) are shown in
Figs.~\ref{fig:tb56fig} - \ref{fig:tb89fig}.  There are well-resolved absorption
bands for most cluster sizes.  The spectra of the smaller sizes, $n=5$, 6, and
7, are dominated by a single intense feature, which may suggest a highly
symmetric structure, while the larger sizes, $n=8$ and $9$ have more richly
featured spectra. For the studied cluster sizes we find absorption bands in the
50--160 cm$^{-1}$ range, which is approximately the range where the optical
phonon branches of the bulk metal were observed by neutron scattering.
\cite{houmann1970lattice}  With few exceptions, all observed metal cluster
vibrations are below the reported value for the stretching frequency of the
neutral Tb$_{2}$ of 137 cm$^{-1}$. \cite{lombardi2002periodic}

To identify the clusters' structures based on the experimental
spectra we perform quantum chemical calculations using density
functional theory (DFT). We use the Perdew-Burke-Ernzerhof (PBE)
functional \cite{perdew1996generalized} for exchange and correlation
as implemented in Turbomole 6.2. \footnote{TURBOMOLE V6.2 2010, a
development of University of Karlsruhe and Forschungszentrum
Karlsruhe GmbH, 1989-2007, TURBOMOLE GmbH, since 2007; available
from {\tt http://www.turbomole.com}.} The treatment of the $4f$
electrons with DFT remains a challenge because they occupy an open
shell and are highly localized to their parent atoms, where
correlation effects are strong.  Their strong localization also
implies that they do not directly participate in chemical bonding. 
Standard DFT functionals are known to falsely delocalize the
$4f$ electrons, and explicit treatment of the $4f$ states requires
corrections such as LDA+U. \cite{dobrich2007electronic}
Alternatively, it is possible to exploit the localization of the $4f$
states and treat them as a part of the core with an effective core
potential.  We follow this approach, using the \emph{ECP54MWB}
effective core potential developed by Dolg et al.  \cite{dolg1989energy,
*dolg1993combination, *yang2005valence}   The valence orbitals are described with
the (6s6p5d)/[4s4p4d] + 2s1p1d (\emph{ECP54MWB-II}) basis set
\footnote{The ECP and basis set are available at: {\tt
http://www.theochem.uni-stuttgart.de/} {\tt pseudopotentials/}}.  
The ECP includes 8 $4f$ electrons in the
core.  This corresponds to the commonly observed trivalent ($4f^8
5d^1 6s^2$) configuration of the terbium atom.

For metal clusters the interrogation of the configurational
space becomes rapidly more complex with growing size and many
locally stable arrangements of atoms, in addition to the true global
minimum, may be identified for any given cluster size. We
investigated a wide range of possible isomers starting our search
with geometries that have been reported for transition metal
clusters (after scaling the interatomic distances to values which
match the bulk values for Tb) including those that have been
identified in our previous studies, \cite{fielicke2007far,
fielicke2004structure, harding2010communications,
harding2010probing} followed by local optimization. In addition, for several
sizes, we used Monte-Carlo basin-hopping sampling
\cite{wales1997global} to locate other local minima. No symmetry
constraints have been applied. For all local minima found the IR
spectra have been calculated analytically based on the harmonic
approximation; the frequencies are not further scaled.  For $n=5-7$,
alternative isomers are typically much higher in energy.
This suggests that the potential-energy surface is quite smooth.

An additional parameter of the calculations is the electronic occupation of the
valence states.  In all of these calculations electronic states with different
numbers of unpaired $5d/6s$ valence electrons have been considered in order to
identify the preferred configuration.  The spin states we refer to in the
following only relate to the $5d/6s$ occupation as the $4f$ electrons are part
of the ECP and not explicitly treated.  If the $4f$ electrons were included, the
total spin of any Tb cluster would be much higher.  (The $4f$ shell of trivalent
Tb has a configuration $J=6,g_L=3/2$ yielding a theoretical moment of $g_L \mu_B
J = 9\mu_B$ per atom. \cite{jensen1991rare}) The ECP calculations also
explicitly neglect any ordering of the $4f$ moments, as well as the $4f-(5d6s)$
exchange interaction, which is known to split the majority and minority ($5d6s$)
spin states for bulk Tb metal in the ferromagnetic state by ~0.1 eV.
\cite{dobrich2007electronic}  An exchange splitting of this magnitude could
plausibly change the ground state valence occupation, and thus the Far-IR
spectra.  The surprising fact is that the $4f$-in-core model, which completely
neglects this effect, still gives remarkably good agreement with the
experimental spectrum.  The frequencies of the IR active bands are correctly
predicted to within 1\%, and the relative intensities match qualitatively.
%
%
%
Figs.~\ref{fig:tb56fig}--\ref{fig:tb89fig} show a comparison between
the experimental far-IR spectra and those of the identified putative
ground states. More isomers and their spectra calculated with a
smaller, presumably less accurate, basis set are reported in
the supporting information \footnote{See Supplemental Material at
[URL will be inserted by publisher] for atomic coordinates,
energies, and calculated IR-spectra of all isomers}. In short, we
can assign the following structures for Tb$_{5-9}^{+}$ (approximate
symmetries are given in parentheses): 5 -- trigonal bipyramid
($C_{2v}$ $2S=4$), 6 -- distorted octahedron (\textbf{6c} -- $D_{2h}$
$2S=1$), 7 -- pentagonal bipyramid or decahedron ($D_{5h}$ $2S=2$),
8 -- bicapped octahedron ($D_{2d}$ $2S=3$), 9 -- tricapped
octahedron ($C_{2v}$ $2S=2$).

For the small clusters, Tb$_{5-7}^{+}$, the best matches to the IR
spectra are obtained with the putative ground states which are calculated to
have 4, 1, and 2 unpaired $5d/6s$ electrons, respectively.  For a given
structure, different spin states are often close in energy, but their calculated IR spectra can differ remarkably, as seen for
Tb$_8^{+}$.  In this case, the doublet state \textbf{8b} is
higher in energy than the quartet \textbf{8a}, but the spectrum of \textbf{8a}
fits the experiment better. The presence of (several) unpaired
$5d/6s$ electrons is consistent with Stern-Gerlach experiments on La
and Y clusters, \cite{knickelbein2005magnetic} which found many
sizes with net magnetic moments ($~0.2$ $\mu_B$/atom).  La and Y
have very similar valence electronic structure to the lanthanides,
but no $4f$ electrons. Thus the observation of high spin in these
clusters supports our finding, even without properly accounting for
the possibility of a $5d6s$ spin polarization induced by the large
localized $4f$ moments.

We find no evidence of multiple isomers for any of these small Tb clusters,
although for Tb$_{9}^{+}$ there are several isomers based on two structural
motifs (\textbf{9a/9b} and \textbf{9c/9d}) which are close in energy (within 0.2
eV) and which have similar spectra. \textbf{9a/9b} is formed by capping three
adjacent faces of an octahedron and \textbf{9c/9d} can be described as two
interpenetrating decahedra.  Both of these could be present, but structure
\textbf{9a/9b} fits the experimental spectra better, and the energy difference
of 0.1 eV is too small to be considered meaningful.

The lowest energy isomer identified by theory does not always give the best
match to the theory, a case illustrated by Tb$_{6}^{+}$.  Tb$_{6}^{+}$ has 17
($5d6s$) electrons, one fewer than the 18 needed to satisfy a spherical shell
closing.  This would indicate a preference for a low-spin distorted geometry
\textbf{6c}, over the octahedral \textbf{6a}. However, the highly symmetric
octahedral structure has high density of states at the HOMO, which increases the
energy gained by exchange splitting in a high spin state.  It is possible for
this effect to stabilize a highly symmetric structure against a Jahn-Teller
distortion.  Indeed, the calculations predict that the $2S=5$ octahedral
structure \textbf{6a} is ~0.3 eV lower in energy than the distorted low-spin
($2S=1$) structure \textbf{6c}, but \textbf{6c} appears to match the
experimental spectrum better.  It is important to note, however, that the
geometric difference between \textbf{6a} and \textbf{6c} is small, (the bond
lengths differ by 2-5\%) and their vibrational spectra are very similar.  The
octahedral structure \textbf{6a} actually has a triply degenerate vibrational
mode located at 100~cm$^{-1}$, which is close to where an intense feature is
observed in the experimental spectrum.  However, the IR intensities calculated
for these modes are very low.  Further investigation is required to understand
whether the presence of the Ar, or the inclusion of the $4f$ electrons would
change the IR intensities of these modes. Thus our experiment cannot determine
conclusively whether Tb$_{6}^{+}$ is a perfect octahedron.


To conclude, we have demonstrated that far-IR spectroscopy allows for structural
assignment for small lanthanide clusters. Our theoretical approach, where the
$4f$ electrons are treated as a part of the core, predicts the frequencies and
IR intensities of the cluster vibrational modes with remarkable accuracy.  This
confirms that, like the bulk materials, the $4f$ electrons remain tightly
localized in metal clusters and do not directly participate in bonding.  The
structures identified are similar to the geometries  found for transition metal
clusters.  In this sense, heavy lanthanides behave like early transition metals
with small $d$-occupation.  It is also highly surprising that the structures can
be determined while the magnetic order and $4f-(5d6s)$ exchange interaction are
completely neglected.  Knowledge of the cluster structures will enable the
application of higher level theory for an explicit treatment of the $4f$
electrons, which in turn will provide a detailed understanding of lanthanide
cluster magnetism.

\begin{acknowledgments}
We acknowledge the support of the Stichting voor Fundamenteel
Onderzoek der Materie (FOM) for providing beam time on FELIX, and
the FELIX staff for their skillful assistance, in particular B.
Redlich and A. F. G. van der Meer. We acknowledge Marek
Sierka, and Pekka Pyykk\"{o} for helpful discussions.  J.B. and D.J.H.
thank the Alexander von Humboldt foundation for financial support.
\end{acknowledgments}


%

\end{document}